\documentclass[prd,twocolumn,showpcs,amsmath,amssymb]{revtex4}

\usepackage{graphicx}
\usepackage{dcolumn}
\usepackage{bm}

\newlength{\picwidth}
\newcommand{\dzero}{D0 } 
\newcommand{\gev}{{\,\rm GeV}}
\newcommand{\pb}{\,\rm{pb}}
\newcommand{\invpb}{\,\rm{pb}^{-1}}

\newcommand{\pythia}{{\sc pythia}} 
\newcommand{\alpgen}{{\sc alpgen}} 
\newcommand{\geant}{{\sc geant}} 

\newcommand{\MvE}{\hat{E}}
\newcommand{\LQ}{\text{LQ}}
\newcommand{\stb} {\sigma  \! \times  \! \beta}
\newcommand{\stbs} {\sigma  \! \times  \! \beta^2}

\newcommand{\totallumiB}{\mbox{$294\pm 19 \invpb$}}

\begin{document}


\hspace{5.2in} \mbox{FERMILAB-PUB-06-455-E}
\title{Search for single production of scalar leptoquarks  
 in {\boldmath $p\bar{p}$} collisions
decaying into  muons and quarks with the \dzero  detector
  \vspace*{0.8cm}
}



%
\author{                                                                      
V.M.~Abazov,$^{35}$                                                           
B.~Abbott,$^{75}$                                                             
M.~Abolins,$^{65}$                                                            
B.S.~Acharya,$^{28}$                                                          
M.~Adams,$^{51}$                                                              
T.~Adams,$^{49}$                                                              
E.~Aguilo,$^{5}$                                                              
S.H.~Ahn,$^{30}$                                                              
M.~Ahsan,$^{59}$                                                              
G.D.~Alexeev,$^{35}$
% ...
\\                                                                            
\vskip 0.30cm                                                                 
\centerline{(D\O\ Collaboration)}                                             
\vskip 0.30cm                                                                 
}                                                                             
\affiliation{                                                                 
\centerline{$^{1}$Universidad de Buenos Aires, Buenos Aires, Argentina}       
\centerline{$^{2}$LAFEX, Centro Brasileiro de Pesquisas F{\'\i}sicas,         
                  Rio de Janeiro, Brazil}                                     
\centerline{$^{3}$Universidade do Estado do Rio de Janeiro,                   
                  Rio de Janeiro, Brazil}                                     
\centerline{$^{4}$Instituto de F\'{\i}sica Te\'orica, Universidade            
                  Estadual Paulista, S\~ao Paulo, Brazil}                     
\centerline{$^{5}$University of Alberta, Edmonton, Alberta, Canada,           
                  Simon Fraser University, Burnaby, British Columbia, Canada,}
\centerline{York University, Toronto, Ontario, Canada, and                    
                  McGill University, Montreal, Quebec, Canada}                
\centerline{$^{6}$University of Science and Technology of China, Hefei,       
                  People's Republic of China}                                 
\centerline{$^{7}$Universidad de los Andes, Bogot\'{a}, Colombia}             
\centerline{$^{8}$Center for Particle Physics, Charles University,            
                  Prague, Czech Republic}                                     
\centerline{$^{9}$Czech Technical University, Prague, Czech Republic}         
\centerline{$^{10}$Center for Particle Physics, Institute of Physics,         
                   Academy of Sciences of the Czech Republic,                 
                   Prague, Czech Republic}                                    
\centerline{$^{11}$Universidad San Francisco de Quito, Quito, Ecuador}        
\centerline{$^{12}$Laboratoire de Physique Corpusculaire, IN2P3-CNRS,         
                   Universit\'e Blaise Pascal, Clermont-Ferrand, France}      
\centerline{$^{13}$Laboratoire de Physique Subatomique et de Cosmologie,      
                   IN2P3-CNRS, Universite de Grenoble 1, Grenoble, France}    
\centerline{$^{14}$CPPM, IN2P3-CNRS, Universit\'e de la M\'editerran\'ee,     
                   Marseille, France}                                         
\centerline{$^{15}$Laboratoire de l'Acc\'el\'erateur Lin\'eaire,              
                   IN2P3-CNRS et Universit\'e Paris-Sud, Orsay, France}       
\centerline{$^{16}$LPNHE, IN2P3-CNRS, Universit\'es Paris VI and VII,         
                   Paris, France}                                             
\centerline{$^{17}$DAPNIA/Service de Physique des Particules, CEA, Saclay,    
                   France}                                                    
\centerline{$^{18}$IPHC, IN2P3-CNRS, Universit\'e Louis Pasteur, Strasbourg,  
                   France, and Universit\'e de Haute Alsace,                  
                   Mulhouse, France}                                          
\centerline{$^{19}$Institut de Physique Nucl\'eaire de Lyon, IN2P3-CNRS,      
                   Universit\'e Claude Bernard, Villeurbanne, France}         
\centerline{$^{20}$III. Physikalisches Institut A, RWTH Aachen,               
                   Aachen, Germany}                                           
\centerline{$^{21}$Physikalisches Institut, Universit{\"a}t Bonn,             
                   Bonn, Germany}                                             
\centerline{$^{22}$Physikalisches Institut, Universit{\"a}t Freiburg,         
                   Freiburg, Germany}                                         
\centerline{$^{23}$Institut f{\"u}r Physik, Universit{\"a}t Mainz,            
                   Mainz, Germany}                                            
\centerline{$^{24}$Ludwig-Maximilians-Universit{\"a}t M{\"u}nchen,            
                   M{\"u}nchen, Germany}                                      
\centerline{$^{25}$Fachbereich Physik, University of Wuppertal,               
                   Wuppertal, Germany}                                        
\centerline{$^{26}$Panjab University, Chandigarh, India}                      
\centerline{$^{27}$Delhi University, Delhi, India}                            
\centerline{$^{28}$Tata Institute of Fundamental Research, Mumbai, India}     
\centerline{$^{29}$University College Dublin, Dublin, Ireland}                
\centerline{$^{30}$Korea Detector Laboratory, Korea University,               
                   Seoul, Korea}                                              
\centerline{$^{31}$SungKyunKwan University, Suwon, Korea}                     
\centerline{$^{32}$CINVESTAV, Mexico City, Mexico}                            
\centerline{$^{33}$FOM-Institute NIKHEF and University of                     
                   Amsterdam/NIKHEF, Amsterdam, The Netherlands}              
\centerline{$^{34}$Radboud University Nijmegen/NIKHEF, Nijmegen, The          
                  Netherlands}                                                
\centerline{$^{35}$Joint Institute for Nuclear Research, Dubna, Russia}       
\centerline{$^{36}$Institute for Theoretical and Experimental Physics,        
                   Moscow, Russia}                                            
\centerline{$^{37}$Moscow State University, Moscow, Russia}                   
\centerline{$^{38}$Institute for High Energy Physics, Protvino, Russia}       
\centerline{$^{39}$Petersburg Nuclear Physics Institute,                      
                   St. Petersburg, Russia}                                    
\centerline{$^{40}$Lund University, Lund, Sweden, Royal Institute of          
                   Technology and Stockholm University, Stockholm,            
                   Sweden, and}                                               
\centerline{Uppsala University, Uppsala, Sweden}                              
\centerline{$^{41}$Physik Institut der Universit{\"a}t Z{\"u}rich,            
                   Z{\"u}rich, Switzerland}                                   
\centerline{$^{42}$Lancaster University, Lancaster, United Kingdom}           
\centerline{$^{43}$Imperial College, London, United Kingdom}                  
\centerline{$^{44}$University of Manchester, Manchester, United Kingdom}      
\centerline{$^{45}$University of Arizona, Tucson, Arizona 85721, USA}         
\centerline{$^{46}$Lawrence Berkeley National Laboratory and University of    
                   California, Berkeley, California 94720, USA}               
\centerline{$^{47}$California State University, Fresno, California 93740, USA}
\centerline{$^{48}$University of California, Riverside, California 92521, USA}
\centerline{$^{49}$Florida State University, Tallahassee, Florida 32306, USA} 
\centerline{$^{50}$Fermi National Accelerator Laboratory,                     
            Batavia, Illinois 60510, USA}                                     
\centerline{$^{51}$University of Illinois at Chicago,                         
            Chicago, Illinois 60607, USA}                                     
\centerline{$^{52}$Northern Illinois University, DeKalb, Illinois 60115, USA} 
\centerline{$^{53}$Northwestern University, Evanston, Illinois 60208, USA}    
\centerline{$^{54}$Indiana University, Bloomington, Indiana 47405, USA}       
\centerline{$^{55}$University of Notre Dame, Notre Dame, Indiana 46556, USA}  
\centerline{$^{56}$Purdue University Calumet, Hammond, Indiana 46323, USA}    
\centerline{$^{57}$Iowa State University, Ames, Iowa 50011, USA}              
\centerline{$^{58}$University of Kansas, Lawrence, Kansas 66045, USA}         
\centerline{$^{59}$Kansas State University, Manhattan, Kansas 66506, USA}     
\centerline{$^{60}$Louisiana Tech University, Ruston, Louisiana 71272, USA}   
\centerline{$^{61}$University of Maryland, College Park, Maryland 20742, USA} 
\centerline{$^{62}$Boston University, Boston, Massachusetts 02215, USA}       
\centerline{$^{63}$Northeastern University, Boston, Massachusetts 02115, USA} 
\centerline{$^{64}$University of Michigan, Ann Arbor, Michigan 48109, USA}    
\centerline{$^{65}$Michigan State University,                                 
            East Lansing, Michigan 48824, USA}                                
\centerline{$^{66}$University of Mississippi,                                 
            University, Mississippi 38677, USA}                               
\centerline{$^{67}$University of Nebraska, Lincoln, Nebraska 68588, USA}      
\centerline{$^{68}$Princeton University, Princeton, New Jersey 08544, USA}    
\centerline{$^{69}$State University of New York, Buffalo, New York 14260, USA}
\centerline{$^{70}$Columbia University, New York, New York 10027, USA}        
\centerline{$^{71}$University of Rochester, Rochester, New York 14627, USA}   
\centerline{$^{72}$State University of New York,                              
            Stony Brook, New York 11794, USA}                                 
\centerline{$^{73}$Brookhaven National Laboratory, Upton, New York 11973, USA}
\centerline{$^{74}$Langston University, Langston, Oklahoma 73050, USA}        
\centerline{$^{75}$University of Oklahoma, Norman, Oklahoma 73019, USA}       
\centerline{$^{76}$Oklahoma State University, Stillwater, Oklahoma 74078, USA}
\centerline{$^{77}$Brown University, Providence, Rhode Island 02912, USA}     
\centerline{$^{78}$University of Texas, Arlington, Texas 76019, USA}          
\centerline{$^{79}$Southern Methodist University, Dallas, Texas 75275, USA}   
\centerline{$^{80}$Rice University, Houston, Texas 77005, USA}                
\centerline{$^{81}$University of Virginia, Charlottesville,                   
            Virginia 22901, USA}                                              
\centerline{$^{82}$University of Washington, Seattle, Washington 98195, USA}  
}                                                                             

\date{December 7,2006}

\begin{abstract}
 We report on a search for second generation leptoquarks ($\LQ_2$) which
decay into a muon plus quark in $p\bar{p}$
 collisions at a center-of-mass energy of $\sqrt{s} = 1.96$ TeV
 in the 
 \dzero detector using an integrated luminosity of about $300 \invpb$. 
   No evidence for a leptoquark signal is observed 
   and an upper bound on the product of the
   cross section for single leptoquark production times branching fraction
    $\beta$  into a quark and a muon was determined for
   second generation scalar leptoquarks as a function of the leptoquark mass.
This result has been combined with a previously published \dzero search for
leptoquark pair production to obtain leptoquark mass limits as
a function of the leptoquark-muon-quark coupling, $\lambda$.
Assuming $\lambda=1$, lower limits on the mass of a second generation scalar leptoquark
coupling to a $u$ quark and a muon
are $m_{ \text{LQ}_2 } > 274 \gev$ and $m_{\text{LQ}_2} > 226 \gev$ for $\beta =1$ and 
$\beta = 1/2$, respectively.

\end{abstract}

\pacs{14.80.-j,13.85.Rm} 

\maketitle

\newpage


The observed symmetry in the spectrum of elementary particles 
between leptons and quarks motivates the existence of leptoquarks
\cite{theory}. Leptoquarks are bosons carrying both quark and lepton 
quantum numbers and fractional electric charge.
 Leptoquarks 
could in principle decay into any combination of a lepton and a quark
that carry the correct charge. 
Experimental limits on lepton number violation, 
on flavor-changing neutral currents, and on proton decay, however, lead to the 
assumption that there would be three different generations of leptoquarks. 
Each of these leptoquark generations couples to only one quark and
one lepton family and, therefore, individually conserves the family lepton numbers \cite{theory2}.
In this Letter, second generation leptoquarks refer to leptoquarks coupling
to muons. 
Since there is no explicit connection between  a given lepton generation 
with any of the three quark generations
 in the standard model,
the second generation leptoquark that couples 
to muons could couple to a quark from any one of the three generations.

\begin{figure}[h]
  \begin{center}
  \mbox{
    \includegraphics[width=.45\picwidth]{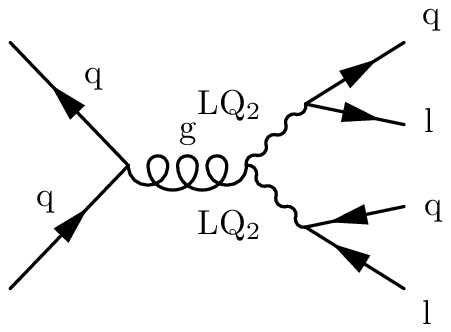}
    \includegraphics[width=.45\picwidth]{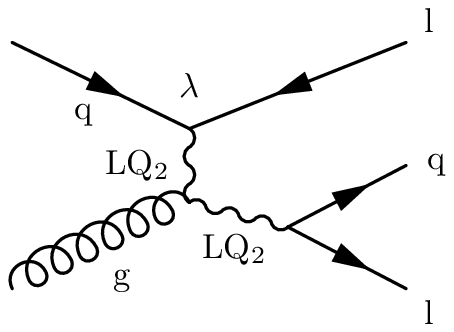}
  }
  \end{center}
  \caption{\label{fig:feynmangraphs} 
    Examples of leading-order Feynman graphs for pair-production (left)
    and single-leptoquark production (right)  of
    leptoquarks.
  }
\end{figure}

Figure \ref{fig:feynmangraphs} shows mechanisms for leptoquark
production and decay in $p\bar{p}$ collisions. 
Leptoquarks can be either pair produced via the strong interaction
or single leptoquark can be produced in association with a lepton.
The cross section
for single leptoquark production depends on the a priori unknown
leptoquark-lepton-quark coupling $\lambda$.
In $p\bar{p}$ collisions, the production cross section for 
single leptoquarks coupling to
up and down quarks is significantly larger 
than that for single leptoquarks coupling to second generation
quarks, and for the search
described in this Letter, we only
considered this scenario. For other quark flavors, the inclusion
of single leptoquark production would  not improve
the sensitivity  from the pair production search even for large couplings.

This search was performed assuming both leptoquark pair and single production
 contribute to the expected signal. 
Therefore both the final state with two jets and two muons  and the final state
with two muons and one jet were considered.
The former has
been studied in previously published  analyses
of leptoquark pair production~\cite{lq-pair,lq-pair-other}.
In addition, limits are given if one assumes that only 
single leptoquark production contributes to the expected signal. 
The  cross section limit for this scenario can be interpreted 
as limit on a final state containing two  energetic muons
and a high $E_T$ jet. 
The inclusion of single leptoquark production in a Tevatron search 
has been previously discussed in Ref.~\cite{single-lq}.

The \dzero detector \cite{run2det}  consists of several layered elements. First is a
magnetic central tracking system which is comprised of a silicon
microstrip tracker (SMT) and a central fiber tracker (CFT), both
located within a 2~T superconducting solenoidal
magnet. Jets are reconstructed from energy depositions in the 
three liquid-argon/uranium
calorimeters: a central section (CC) covering pseudorapidities, $\eta=- \ln [\tan( \theta/2)]$, 
where $\theta$ is the polar angle with respect to the proton beam direction,
up to $|\eta| \approx 1$, and two endcap calorimeters (EC) extending coverage to $|\eta|\approx
4$, all housed in separate cryostats~\cite{run1det}. 
Scintillators between the CC and EC cryostats
provide sampling of developing showers at $1.1<|\eta|<1.4$. 
A muon system  \cite{muonNIM} resides beyond the calorimetry and consists of a layer
of proportional wire tracking detectors and scintillation trigger 
counters before $1.8\,\rm{T}$
toroids, followed by two similar layers after the toroids. The 
 muon system is used for triggering and identifying muons.
 The muon momenta are measured from the
curvature of the muon tracks in the central tracking system. 

The data used in this analysis were collected between August 2002 and
July 2004, corresponding to an  
integrated luminosity of \totallumiB{}. Only events
which pass single- or di-muon triggers were
considered. At the first trigger level, a muon was triggered by a
coincidence of hits in at least two of the three scintillator 
layers of the muon system within a time window consistent with muons
coming from the interaction point. At the second trigger level, a muon
track was identified from the hits in the  proportional wire tracking detectors and
the scintillators of the muon system.
The overall trigger efficiency  for $\mu j+\mu j$ and $\mu +\mu j$ 
events fulfilling the selection criteria of this analysis 
was measured to be $(89\pm3)\%$. 

Muons in the region $|\eta|<1.9$ were reconstructed from hits 
in the three layers of the muon
system which could be matched to isolated tracks in the
central tracking system. Cosmic muon events were rejected by cuts on
the timing in the muon scintillators and by removing back-to-back
muons. Jets were reconstructed using the iterative midpoint cone 
algorithm \cite{conealgo} with a cone size of $0.5$. The jet energies
were calibrated as a function of the jet transverse energy ($E_T$) and $\eta$
by imposing transverse energy balance in photon-plus-jet events. Only jets which were
well-contained within the detector were considered by requiring 
$|\eta|<2.4$.

For this search, the background is dominated by 
Drell-Yan production and $Z$ decays:
$Z/\gamma^*\rightarrow\mu\mu$ (+jets)  ($Z$/DY).
Additional backgrounds coming from QCD multijet production and from $W$+jets events
(with at least one reconstructed muon originating not
from the hard scattering) were estimated and found to be negligible.
 To evaluate the
contribution from the $Z$/DY background, samples of Monte Carlo (MC) events
were generated with \pythia{} (Version 6.202) \cite{pythia}.
Samples of  $t\bar{t}$ ($m_t=175\gev$) and
 $WW$ production were also generated with \pythia{}.
The signal efficiencies were calculated using \pythia{} samples of
 ${\text{LQ}_2}+\mu\rightarrow\mu j+\mu$ and 
${\text{LQ}_2}\overline{\text{LQ}_2} \rightarrow\mu j+\mu j$ MC
events for leptoquark masses ($m_{\text{LQ}_2}$) from $140$ to $280\gev$ in steps of $20\gev$.
All MC events were processed using a full simulation of the
\dzero detector based on \geant{} \cite{geant} and the standard
event reconstruction.
Differences in the trigger and reconstruction efficiencies between data
and Monte Carlo were taken into account using proper weightings of the
MC events. 

The search for  leptoquarks required two muons with transverse momenta
$p_T > 15\gev$ and either one or two jets with $E_T^j >25\gev$. 
To reduce the $Z$/DY background at high dimuon mass
due to occasionally poorly reconstructed muon tracks, 
 advantage was taken of the fact
 that no or little missing transverse energy
is expected in either signal or $Z$/DY events. The missing
transverse energy was determined from the transverse energy imbalance of all
muons and jets ($E_T>20\gev$) in the event, and the momentum of the
muon opposite to the direction of the missing transverse energy
({\rm i.e.} with the larger azimuthal angle relative to the direction
 of the missing  transverse energy)
was corrected such that the missing transverse energy
parallel to the muon vanished. 
This degraded the muon momentum resolution
   and shifted the dimuon mass to lower values in both data and MC.
   However, this correction allowed the suppression of poorly reconstructed
   $Z$ boson events shifted into the high mass region where the search for leptoquarks 
   was taking  place.

\begin{figure}[!!!t]
\includegraphics[width=0.9\picwidth]{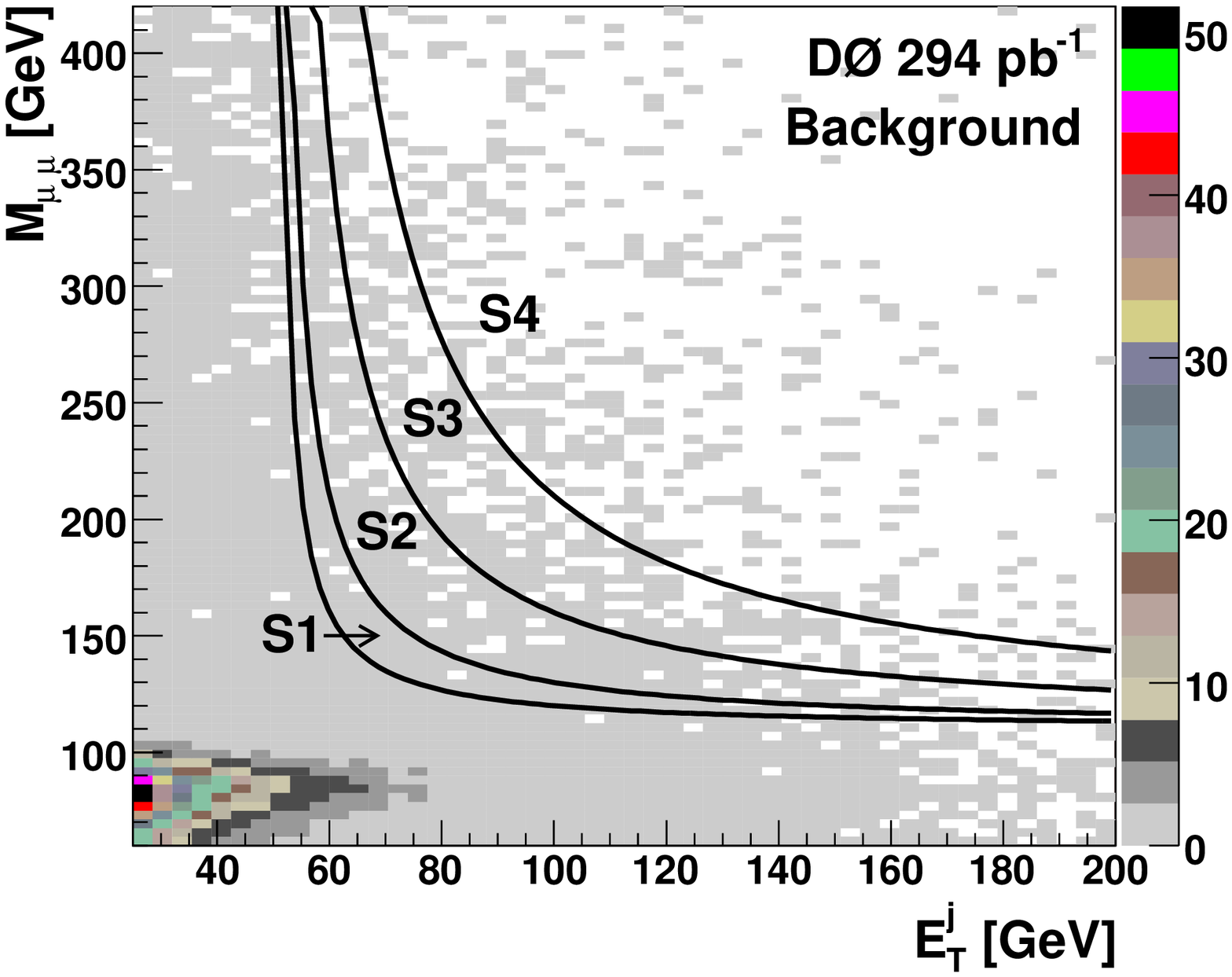}
\includegraphics[width=0.9\picwidth]{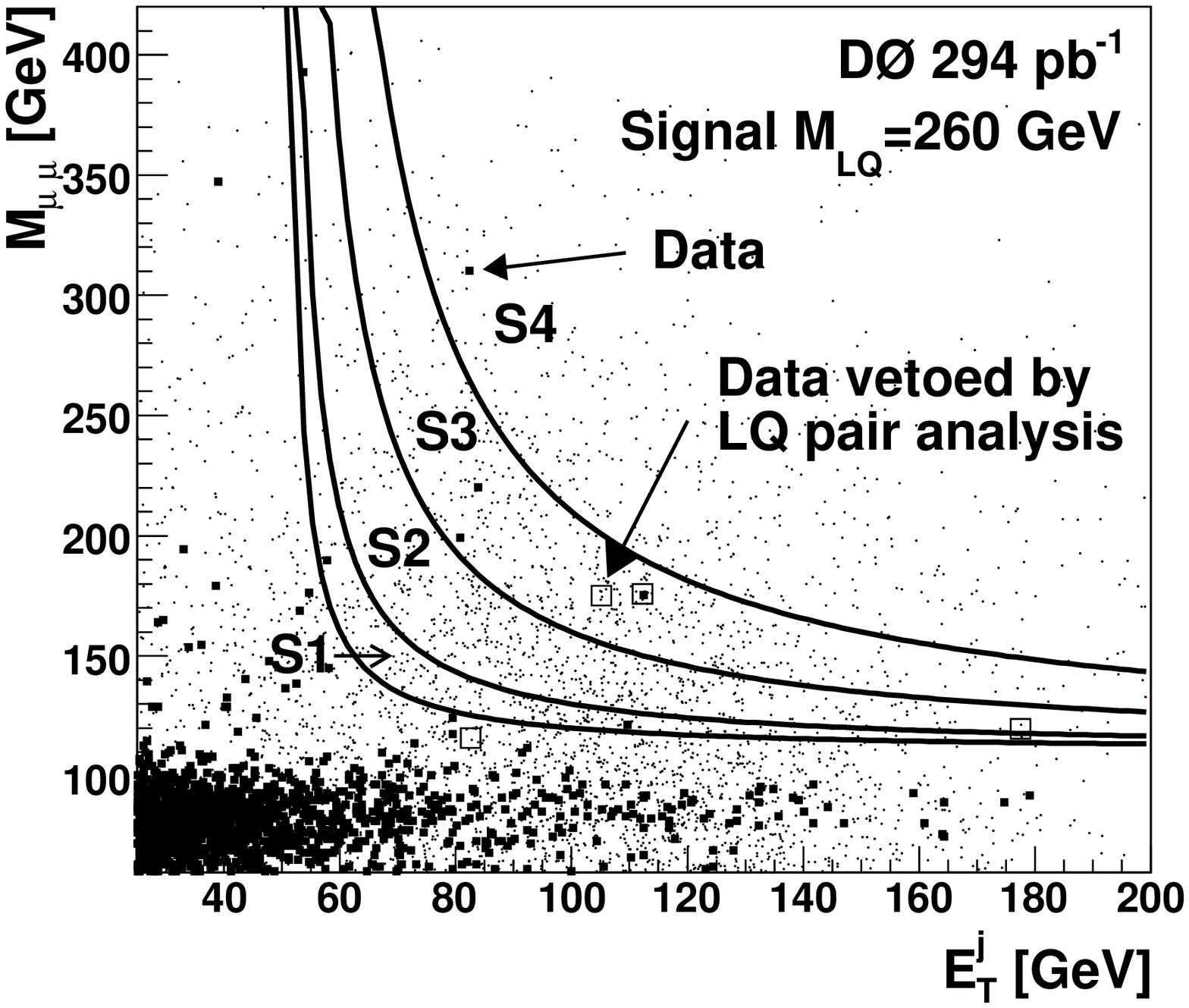}
\vspace*{-0.4cm} 
\caption{\label{fig:ETvM}  
 Distributions of $M_{\mu\mu}$ vs. the highest jet $E_T^{j}$ for
background (upper plot) and signal and data (lower plot).
The curved lines show the edges of the signal bins (see text for definition).  
The bottom plot shows as small dots the expected signal 
distribution from single leptoquark production ($m_{\text{LQ}_2}=260\gev$). 
The bottom plot also shows the
observed data for events  not classified as leptoquark pair candidates 
(small black boxes)
and  classified as leptoquark pair candidates (open boxes).
}
\end{figure}

To create statistically independent signal bins, events were first classified
as either leptoquark pair or single leptoquark  candidates.
Events were classified as leptoquark pair candidates if they contained two
jets with $E_T>25\gev$, had a dimuon mass $M_{\mu\mu}>105\gev$ (to remove $Z$ boson events), 
and  fulfilled the requirement
$\hat{S}=S_T/\gev - 0.003\times (M_{\mu\mu}/\gev-250)^2 > 250$, with $S_T$ defined 
as the sum the 
sum of the absolute values of the transverse energies of the two jets and the transverse 
momenta of the two muons forming the $\mu j+\mu j$ system.
Events not classified as leptoquark pair candidates were classified
as single leptoquark candidates if they contained at least one jet with
$E_T>50\gev$,  had a di-muon mass $M_{\mu\mu}>110 \gev$ and fulfilled
the requirement $\MvE=(M_{\mu\mu}/\gev-110) \times (E_T^{j}/\gev-50) > 500$
(see Fig.\ \ref{fig:ETvM}).
The optimum choice of variables and cut
values has been determined to optimize the sensitivity to the signal. These
selections have been cross-checked with a neural net optimization, which
gave similar results.
Eleven events were either classified as  leptoquark pair candidates
or  single leptoquark candidates
while  $6.6\pm0.5\text{ (stat)}\pm1.1\text{ (syst)}$ are expected
from standard model background.
A small excess of data over background was observed.
The probability that $6.6 \pm 1.2$  expected events fluctuate up to
11 observed events is $9.2\%$.

Candidate events were arranged in  bins of
increasing signal to background ratio. 
For leptoquark pair candidates, bin boundaries
of  $\hat{S}=320$ and $390$   are used to
define  bins P1, P2 and P3 \cite{lq-pair}. For single leptoquark candidates, boundaries
of $\MvE = 1000,2500,5000$
were used to define signal bins S1, S2, S3 and S4 (Fig.\ \ref{fig:ETvM}).

\begin{table*}[!!!t] 
\caption{\label{tab:cuflow}
  Signal efficiency ($\varepsilon_{\text{single} }$) for
  selecting single leptoquarks for $\beta=1$,
  number of data events ($N_{\text{data} }$),
   and number of predicted background events
  ($N_{\text{pred}}^{\text{bgd}}$). The first uncertainty on $N_{\text{pred}}^{\text{bgd}}$
  is due to limited Monte Carlo statistics, the second denotes the
  systematic uncertainty. The first two lines indicate the total
  number of events after the initial event selection while the other lines indicate the
  numbers for the individual bins of the leptoquark pair candidates (P1--P3)
and single leptoquark candidates (S1--S4) as described in the text.
}
\begin{ruledtabular}
\begin{tabular}{ccccc}
  Cut
  & $\varepsilon_{ \text{single} }$
  & $\varepsilon_{ \text{single} }$
  & $N_{\text{data}}$
  & $N_{\text{pred}}^{\text{bgd}}$
  \\
  & $(m_{\text{LQ}_2}=200\gev)$
  & $(m_{\text{LQ}_2}=240\gev)$
  & 
  & 
  \\
  \hline
$ M_{\mu\mu}>110 \gev$  &
      $ 0.145 \pm 0.013 $ &  $ 0.176 \pm 0.015 $   &$43$ & $44.75 \pm 1.74 \pm 6.13$  \\
 $ E_T^{j}>25 \gev$ & &  & &    \\  \hline
$ M_{\mu\mu}>110 \gev$ &
      $ 0.122 \pm 0.012 $ &  $ 0.158 \pm 0.014 $   &$20$ & $13.41 \pm 0.92 \pm 1.57$  \\
$ E_T^{j}>50 \gev$   & &  & &   \\  \hline 
P1 &  $ 0.011 \pm 0.002 $ &  $ 0.015 \pm 0.002 $   & $2$ & $ 0.96 \pm 0.25 \pm 0.28$  \\
P2 &  $ 0.006 \pm 0.001 $ &  $ 0.011 \pm 0.002 $   & $2$ & $ 0.39 \pm 0.10 \pm 0.11$  \\
P3 &  $ 0.006 \pm 0.001 $ &  $ 0.012 \pm 0.002 $   & $0$ & $ 0.27 \pm 0.10 \pm 0.08$  \\
S1 &  $ 0.018 \pm 0.002 $ &  $ 0.014 \pm 0.002 $   & $2$ & $ 2.01 \pm 0.33 \pm 0.57$  \\
S2 &  $ 0.028 \pm 0.003 $ &  $ 0.030 \pm 0.003 $   & $1$ & $ 1.61 \pm 0.27 \pm 0.44$  \\
S3 &  $ 0.016 \pm 0.002 $ &  $ 0.029 \pm 0.003 $   & $3$ & $ 0.87 \pm 0.17 \pm 0.29$  \\
S4 &  $ 0.015 \pm 0.002 $ &  $ 0.029 \pm 0.004 $   & $1$ & $ 0.44 \pm 0.08 \pm 0.06$  \\ \hline   
Signal bins (P1-P3,S1-S4)
   &  $ 0.100\pm 0.010  $ &  $ 0.140 \pm 0.013 $   &$11$ & $ 6.55 \pm 0.53 \pm 1.08$  \\  
\end{tabular}
\end{ruledtabular}
\end{table*} 

Table \ref{tab:cuflow} summarizes 
the efficiency of the single leptoquark selection for two leptoquark masses
 as well as the number of expected
background events and the distribution of the data in the 
three pair and four single leptoquark bins.

The dominant uncertainties on the predicted number of background
events are Monte Carlo statistics, varying between $7\%$ and $25\%$
for the seven signal bins, the jet-energy scale uncertainty [$(2$ -- $12)\%$],
and the  uncertainty on the  shape of the jet transverse energy distribution 
of $Z$/DY events [$(20$ -- $30)\%$].  
The latter has been estimated by a comparison of the \pythia{}
\cite{pythia} simulation
with Monte Carlo events generated with the \alpgen{} \cite{alpgen} 
event generator.
For leptoquark pair candidates, the uncertainty due to the two jet 
requirement was estimated to be $16\%$~\cite{lq-pair}. 
In addition, the following sources of systematic  uncertainties
were taken into account: luminosity ($6.5\%$), theoretical cross
section of the $Z$/DY processes ($3.6\%$), and muon triggering and 
identification ($5\%$). The systematic uncertainties, added in quadrature, 
are shown in Table\ \ref{tab:cuflow}.

The systematic uncertainties on the signal efficiencies arise from 
the jet-energy scale uncertainty [$(3$ -- $17)\%$], 
muon triggering and identification ($5\%$),
limited Monte Carlo statistics [$(4$ -- $14)\%$, uncorrelated between bins], 
and  parton distribution function uncertainty ($2\%$).

Limits on the cross section for single leptoquark production were
calculated from the observed and expected events in the seven
signal bins (P1-P3 and S1-S4). Three different scenarios were considered: 
(a) the only contribution to the signal region came
from standard model background  and from single leptoquark production,
(b) contributions to the signal region came from 
standard model background  and single 
leptoquark production plus leptoquark pair production
 corresponding to 
the nominal leptoquark pair cross section~\cite{kraemer1} with
$\beta=BR({\text LQ} \rightarrow \mu q)= 1/2$, and (c) as (b)
but with $\beta=1$.
For scenarios (b) and (c) leptoquark pair events in the signal
bins P1-P3 are treated exactly the same way as in~\cite{lq-pair}.
The analysis described above can therefore be considered as a combination 
of the search for singly produced  leptoquark  with the leptoquark
pair analysis published in~\cite{lq-pair}.

\begin{figure}
  \begin{center}
  \mbox{
    \includegraphics[width=0.95\picwidth]{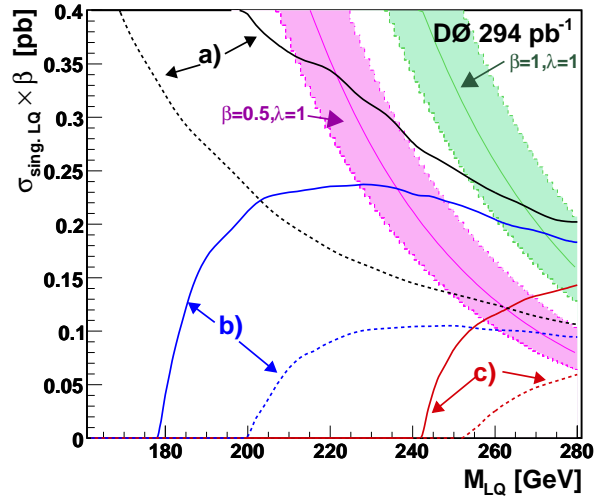}
  }
  \end{center}
  \vspace*{-0.4cm}
  \caption{\label{fig:lim_cross}
    Cross section upper limits at 95\% C.L. for the production of single leptoquarks 
for the three scenarios (a) no contribution from
leptoquark pair production, (b)  pairs contribute with a $\stbs$ corresponding
to $\beta=1/2$, and (c)  pairs contribute with a $\stbs$ corresponding
to $\beta=1$.
    The solid line is the observed limit and the dashed line the average
    expected limit assuming that no signal is present.
    Also indicated are the predicted single leptoquark production 
     cross sections for $\lambda=1$, $\beta=1$ 
     and $\lambda=1$, $\beta=1/2$. The shaded region is the variation 
     of the cross section using renormalization and factorization scales
     of $2 \times m_{\text LQ_2}$ and $0.5 \times m_{\text LQ_2}$, respectively.
  }
\end{figure}

These calculations were performed assuming a flat prior and Gaussian
errors as described in Ref.~\cite{tomjunk}
with the correlations between the systematic errors taken into account. 
The observed limit was calculated using the confidence level 
ratio~\cite{tomjunk}  $CL_S =
CL_{S+B}/CL_{B}$, where $CL_{S+B}$ is the confidence level
for the signal plus background hypothesis, and $CL_{B}$ is the
confidence level for the background only.

Figure \ref{fig:lim_cross} shows the 95\% C.L.
exclusion limits for the production cross section times branching fraction
for single 
leptoquarks as functions of the leptoquark mass.
The predicted cross section for single leptoquarks depends both on
the coupling strength $\lambda$ and on the leptoquark mass.
Each point in Fig.\ \ref{fig:lim_cross} therefore corresponds to a
specific  value of  $\lambda$ and branching fraction. 
In scenarios (b) and (c), leptoquark masses are excluded independently
of the single leptoquark cross section if they 
are already excluded because of the leptoquark pair production results.
The average expected limits  are better than the observed ones because
of the small excess of data over background. 

\begin{figure}[!!!t]
  \begin{center}
  \mbox{
    \includegraphics[width=\picwidth]{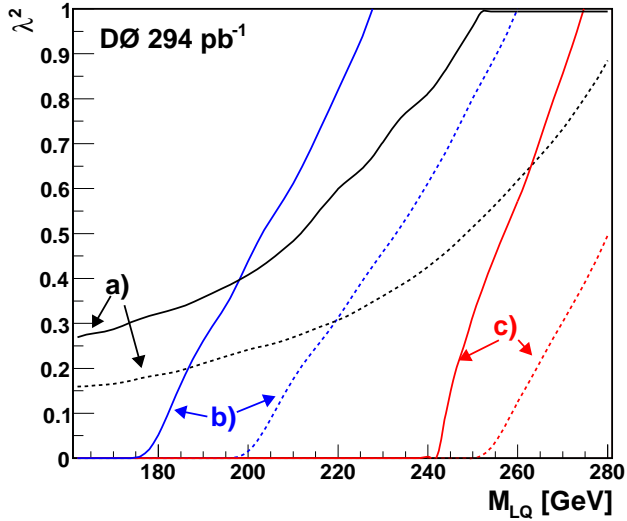}
  }
  \end{center}
  \vspace*{-.3cm}
  \caption{\label{fig:lim_lam}
    Upper limits on $\lambda^2$ for the three scenarios: (a) no contribution from
leptoquark pairs events and $\beta=1$,  (b) leptoquark pairs contribute with 
a $\stbs$ corresponding to $\beta=1/2$, and (c) leptoquark pairs contribute 
with a $\stbs$ corresponding to
$\beta=1$.
The solid line is the observed and the dashed line is the expected limit.
The regions above the solid lines are excluded at 95\% C.L. 
  }
\end{figure}

Also indicated are the predicted leading order cross sections for
single leptoquarks \cite{single-lq-cross} for $\lambda=1$ 
with $\beta=1$ and $\beta=1/2$.
For the production cross sections, it was assumed that the leptoquark couples to
the $u$ quark and a muon.
From the intersection of the cross section limit with the lower limit of the predicted
cross section, which were derived by using renormalization and factorization scales
of $2 \times m_{\text LQ_2}$, respectively, 
exclusion limits on the leptoquark mass as a function of the coupling
strength $\lambda$ can be estimated.
Figure \ref{fig:lim_lam} shows the 95\% C.L.
exclusion regions as functions of the leptoquark mass and $\lambda^2 $.
For scenario (b), the reduction of expected events due to $\beta=1/2$ applies
to both single
and  pair production of leptoquarks.
The intersection of the curves with $\lambda^2=0$ yields the result when only
leptoquark pairs are considered.

The production of a leptoquark-like particle  in association with a muon 
which  decays into a jet and a muon can be excluded for cross section times branching fractions
$ \stb > 0.40~\pb$ for a particle mass of $200 \gev$ and 
$\stb > 0.23~\pb$ for a particle mass of $260 \gev$.
The  \dzero Run II result for scalar leptoquarks and $\beta=1$ of 
$m_{\text LQ_2}(\lambda^2 \ll 1) > 247\gev$, which
only considered leptoquark pair production~\cite{lq-pair}, is improved for an
assumned  leptoquark coupling
to  a $u$ quark and a muon of
$\lambda^2=1$ to  $m_{\text LQ_2}(\lambda^2=1) > 274~\gev$. 
For $\beta=1/2$, the improvement is from
$m_{\text LQ_2}(\lambda^2 \ll 1) > 190\gev$ to  $m_{\text LQ_2}(\lambda^2=1) > 226 \gev$.
For $\lambda^2=0.1$ the observed limits show no improvement while
the expect limits increase by about $7 \gev$. 
This analysis is the first search for single leptoquark production in  $p\bar{p}$ collisions.


\begin{acknowledgments}
%
We thank the staffs at Fermilab and collaborating institutions, 
and acknowledge support from the 
DOE and NSF (USA);
CEA and CNRS/IN2P3 (France);
FASI, Rosatom and RFBR (Russia);
CAPES, CNPq, FAPERJ, FAPESP and FUNDUNESP (Brazil);
DAE and DST (India);
Colciencias (Colombia);
CONACyT (Mexico);
KRF and KOSEF (Korea);
CONICET and UBACyT (Argentina);
FOM (The Netherlands);
PPARC (United Kingdom);
MSMT (Czech Republic);
CRC Program, CFI, NSERC and WestGrid Project (Canada);
BMBF and DFG (Germany);
SFI (Ireland);
The Swedish Research Council (Sweden);
Research Corporation;
Alexander von Humboldt Foundation;
and the Marie Curie Program.

\end{acknowledgments}

\end{document}